\documentstyle[multicol,epsf,aps]{revtex}
\def\ruleleft{\vspace{-2.5\baselineskip}\begin{multicols}{2}\ \linebreak\vspace{-\baselineskip}\hrulefill\raisebox
{0.84mm}{$\!\rfloor$}\[\]\end{multicols}\vspace{-1.5\baselineskip}}
\def\ruleright{\vspace{-1.5\baselineskip}\begin{multicols}{2}\ \linebreak\raisebox
{-2.45mm}{$\lceil\!$}\hrulefill\end{multicols}\vspace{-\baselineskip}}
\newlength{\normalarraycolsep}
\newlength{\normaltabcolsep}
\newcommand{\be}{\begin{equation}}
\newcommand{\ee}{\ \ \end{equation}}
\newcommand{\bea}{\setlength{\arraycolsep}{0.4\normalarraycolsep}
                  \begin{eqnarray}} 
\newcommand{\eea}{\ \ \end{eqnarray}\setlength{\arraycolsep}
                  {\normalarraycolsep}}
\newcommand{\bean}{\setlength{\arraycolsep}{0.4\normalarraycolsep} 
                  \begin{eqnarray*}}
\newcommand{\eean}{\ \ \ \end{eqnarray*}\setlength{\arraycolsep}
                  {\normalarraycolsep}}

\begin{document}
\title{\bf Liquid-state theory of charged colloids}
\author{\bf Yan Levin, Marcia  C. Barbosa and M. N. Tamashiro}
\address{\it 
Instituto de F\'{\i}sica, Universidade Federal
do Rio Grande do Sul\\ Caixa Postal 15051, 91501-970
Porto Alegre (RS), Brazil\\
{\small levin@if.ufrgs.br, barbosa@if.ufrgs.br, mtamash@if.ufrgs.br}}
\maketitle
\begin{abstract}
A simple theory of the fluid state of a  
charged colloidal suspension is proposed. The full free
energy of a polyelectrolyte solution is calculated.
It is found that the 
counterions condense onto
the polyions forming clusters 
composed of one
polyion and $n$ counterions. The distribution of cluster
sizes is determined explicitly.  In agreement with the current experimental 
and Monte Carlo results, {\it no}\/ liquid-gas phase separation was
encountered.
\end{abstract}
\bigskip

\centerline{{\bf PACS numbers:} 05.70.Ce; 61.20.Qg; 61.25.Hq}

\begin{multicols}{2}

The thermodynamic properties of systems in which
the predominant  interactions are due to the long-ranged 
Coulomb potential still remain largely
not understood in spite of the tremendous effort that
has been invested over the span of this century.
 Nevertheless, it would be
unfair to say that   no great progress has been 
done. The pioneering work of Debye and H\"uckel \cite{DH23}
has lead to our understanding  of dilute electrolyte
solutions. The subsequent improvements by Bjerrum extended
the validity of the limiting laws to larger densities \cite{Bj26}. These
developments were followed by the introduction of powerful
integral equations and by the computational methods such
as Monte Carlo {\it (MC)}\/
simulations \cite{Ch73}.
Surprisingly,
the theoretically obtained coexistence curve \cite{Fi93},
that is in closest
agreement with {\it MC,} is based on the fundamental ideas
advanced by Debye, H\"uckel and Bjerrum 
more than $70$ years ago \cite{DH23,Bj26}. 
The simplicity and the transparency of the ideas forming the basis
of 
the Debye-H\"uckel-Bjerrum  {\it (DHBj)}\/ theory  makes
it easy to apply to other coulombic systems \cite{Le94}.

The charged colloidal suspensions present a 
severe challenge to any statistical
mechanics theory. The asymmetry between the charge on a
polyion and a counterion, which can be as high as 10000:1, makes
the
usual integral equations of the liquid-state theory impossible to
solve.
For low charge asymmetry, less than 20:1, it was found that there
exists a
region in the temperature density plane where the Hyperneted
Chain equation {\it (HNC)}\/ ceases to have
solutions \cite{Be85}. This could be interpreted
as a region of instability, in which the sample phase separates into the
coexisting liquid and gas.
It is, however, still unknown to what extent the break down
in {\it HNC}\/ equation can be attributed to the
underlying phase separation, since the region of instability of
{\it HNC}\/ does not coincide with the true spinodal line
\cite{Fi81}. Furthermore,
the extensive experimental and simulation search for this
gas-liquid transition for polyelectrolytes has, so far, proven
to be futile \cite{Vl89}.

At high volume fractions the strongly charged polyions form a
lattice
({\it bcc}\/ or {\it fcc}). This ``solid state" provides us with a
major
simplification in that each polyion can be studied individually 
inclosed in its own Wigner-Seitz cell and  surrounded by its
own counterions \cite{Al84}. Unfortunately, once 
the lattice 
melts the cell picture is no longer valid \cite{Lo93}. As  is
usual, the
liquid state is
significantly more complex than the solid state.

In this letter we shall attempt to construct a theory for 
the fluid state of a polyelectrolyte solution.
We shall work in the context of the Primitive Model of Polyelectrolyte
{\it (PMP)}\/ \cite{Le94}. Our system  will consist of $N_{p}$ polyions
 inside
a volume $V$. The polyions will be idealized as hard
spheres of radius $a,$ each carrying $Z$ ionized  groups
of charge $-q$ uniformly spaced along the surface. A total
of $ZN_{p}$ counterions will be present  to preserve the overall charge
neutrality of
the system. For simplicity, we shall take the counterions to be
point-like 
and to carry charge $+q$. The solvent will be represented
as a uniform medium of a dielectric constant $D$.
As was pointed out by Onsager \cite{On33}, the
full nonlinear Poisson-Boltzmann {\it (PB)}\/ equation is 
electrostatically inconsistent for the asymmetric systems,
linearization then, besides
simplifying 
the calculations, is an important step in restoring the
self-consistency
of the theory. The fundamental assumption behind the {\it DHBj}\/ 
theory is that the nonlinearities omitted in the process of
linearization
of the {\it PB}\/ equation can be reintroduced into the theory
through
the allowance for ion association.  In general, we expect that 
the fluid state of the asymmetric electrolyte will be composed of
free unassociated polyions of density $\rho_{0}$,
free counterions of density $\rho_{f}$, and of clusters
consisting of {\it one}\/
polyion and $0< n \leq Z$
associated counterions. The density of
clusters with $n$ counterions is $\rho_n$. In the discussion that
will follow we shall suppose that the condensed counterion neutralizes
one of the polyion charges,
in such a way that the effective surface charge of a
$n$-cluster is $\sigma_n=-q(Z-n)/(4\pi a^2)$. It is evident that
 $\rho_p=\sum_{n=0}^{Z} \rho_{n}$ and 
$Z\rho_{p}=\rho_{f}+\sum_{n=0}^{Z} 
n\rho_{n}$, where $\rho_p=N_{p}/V$, is
the total density of polyions (associated or not). All the
thermodynamic properties of the polyelectrolyte solution can be determined
once the free energy is known. In particular the osmotic pressure is a Legendre
transform of the Helmholtz free-energy density, $f=-F/V$, 
$p=f(T,\{\rho_s\})+\sum_{s}^{} \mu_s\rho_s$, where the chemical potential of a specie
$s\in\{f,n\}$ is $\mu_s=-\partial f/\partial \rho_s$.
The free energy can be expressed as a sum of electrostatic and entropic 
contributions. The electrostatic free energy is due to 
{\it the polyion-counterion, the polyion-polyion,}  and
{\it the counterion-counterion}\/ interactions.  
The {\it polyion-counterion}\/ contribution
 can be obtained in the framework of
the usual {\it DH}\/ theory.

Let us concentrate our attention 
on one cluster of size $n$ fixed at $r=0$.
Due to  
the excluded volume, no counterions will
be found inside $r<a$. Therefore the electrostatic
potential $\Psi_n$  in this region satisfies
the Laplace 
equation $\nabla^2 \Psi_n=0$.
Outside $r> a-\epsilon$, the mean charge distribution
will be specified by the cluster-counterion 
correlation function. Within the 
{\it DH}\/ theory this is approximated
by a Boltzmann factor  leading 
to the charge density 
$\rho_{q}(r)=-\sum_{n=0}^{Z}q(Z-n)\rho_{n}+q\rho_{f}
e^{-\beta q\Psi_n (r)}+\sigma_n\delta(|\vec{r}|-a)$.
Notice that only  free 
unassociated  counterions get polarized; the free
unassociated polyions and clusters
are too massive to be affected by the
electrostatic fluctuations and only contribute
to the neutralizing background. Substituting
this expression  into the Poisson equation,
$\nabla^2 \Psi_n=-4\pi\rho_{q}/D$,
one obtains 
the nonlinear Poisson-Boltzmann equation.
After the linearization of  the exponential factor,
we are lead to the Helmholtz equation,
$\nabla^2 \Psi_n=\kappa^2\Psi_n$,
where $\kappa a=\sqrt{4\pi \rho_{f}^{\ast}/T^{\ast}}$
and the reduced temperature and density 
are respectively
$T^{\ast}\equiv
aDk_{B}T/q^2$                and
$\rho^{\ast}=\rho a^3$. In principle the linearization is
valid only in the limit $\beta q \Psi_n\ll 1$, however, when
the formation of clusters is properly taken into account
the validity of the theory extends far into the nonlinear regime
\cite{Fi93}.

Both the Laplace and the Helmholtz equations can now be solved,
supplemented
by 
the boundary condition of continuity of the electrostatic
potential,
and  discontinuity  in the normal component of  the electric
field related to the presence of surface charge at $r=a$
\cite{Text}.
Under these conditions,
it is easily found that the 
electrostatic  potential of a $n$-cluster is 
$\Psi_n^{\rm in}(qn,qZ)=-(Z-n)q/(Da(1+\kappa a))$
for $r< a$, and
$\Psi_n^{\rm out}(qn,qZ)=-(Z-n)q e^{\kappa (a -r)}/(Dr(1+\kappa a))$ for 
$ r\geq a$.
The electrostatic energy of a $n$-cluster
is
\be
\label{U}
\beta U_n (q n,qZ)
= 2\pi \beta  \int_{a}^{\infty}\rho_{q}(r) \Psi_n(q n,qZ)
r^2 dr \:,                               
\ee
where $\beta=1/(k_{B}T)$.
The electrostatic {\it free-energy
density}\/ for the polyion-counterion interaction is obtained through the 
Debye charging process, where all 
the particles are charged from 0 to their final charge
\cite{DH23,Ma55},
\bea
\label{FDH}
\beta f_{DH}&=&-\sum_{n=0}^{Z}\rho_n\int_{0}^{1} 
\frac{2\beta U_n (\lambda q n,\lambda qZ)}{\lambda}     
d\lambda \nonumber \\
&=& -\sum_{n=0}^{Z}   \frac{(Z-n)^2}{2T^{\ast}(1+\kappa a)}\rho_n \:.  
\eea

The polyion-polyion contribution to the free energy can be
calculated
in the spirit of the usual Van-der-Waals {\it (VdW)}\/ theory \cite{Fi93}. In
particular, the tremendous difference in mass between the
polyion and the counterion, $m_{c}/m_{p}\ll 1$, leads to the 
effective separation in characteristic
time scales $\tau_{c}\ll\tau_{p}$.
Thus, any change 
in the configuration of one
polyion will be accompanied by an almost instantaneous
rearrangement of the counterion cloud.
Under these conditions, the degrees of freedom associated with the
motion of counterions can be effectively integrated out, resulting
in a {\it short}-ranged
effective pair potential of a {\it DLVO}\/ form \cite{DLVO},
$V_{n,m}^{\rm eff}=q^2(Z-n)(Z-m)[\theta(\kappa a )]^2e^{-\kappa r}/(Dr)$,
where the enhancement factor $\theta(\kappa a)=e^{\kappa a}/(1+\kappa a)$ is
the result of the absence of screening inside the volume
occupied by two polyions. Within the {\it VdW}\/ theory
\end{multicols}
\ruleleft
\medskip

\be
\label{FPP}
\beta f_{PP} = -\sum_{n=0,m=0}^{Z,Z}
\frac{\beta \rho_n\rho_{m}}{2}\int_{}^{} V_{n,m}^{\rm eff}(r)\,
d^3r  = -\sum_{n=0, m=0}^{Z,Z}
\frac{2\pi a^3 (Z-n)(Z-m)\rho_n\rho_{m}}{T^{\ast}}\;
\frac{1+2\kappa a}{(\kappa a)^2 (1+\kappa a)^2} \;  .
\ee 

\ruleright
\begin{multicols}{2}
Now, the electrostatic free energy 
due to the
interactions between
free ions can be estimated as that of a
One Component Plasma {\it (OCP)}\/ \cite{Ab59}. In particular, although
within our treatment the counterions are taken to be  point particles,
the electrostatic repulsion
between the two equally charged counterions will keep them from
approaching each other closer than a distance $d$.
This can be calculated using the {\it OCP}\/ theory
to be $d=[(1+3\kappa a /T^{\ast})^{1/3}-1]/\kappa$ [16(a)].
As one might have expected, in the limit of
small densities this reduces to the Bjerrum 
length, $d\approx a/T^{\ast}\equiv \lambda_B$.
The electrostatic free energy
is found through the Debye charging process and an analytic expression valid over a wide range of coupling strengths is presented in Ref.~[16(b)].

Finally, the entropic (mixing) contribution to the free energy is 
expressed using the Flory theory as
$\beta f_{EN}=\sum_{s}^{} [ \rho_s -\rho_s\ln (\phi_s/\zeta_s)]$,
where $\phi_s$ is the volume fraction occupied by each specie $s$;
$\phi_n=4\pi\rho_n a^3/3$,
 $\phi_{f}\equiv 4\pi\rho_f d^3/3$, and $\zeta_s$ is the internal partition
 function of $s$, $\zeta_0=\zeta_f=1$,  
and  $\zeta_n=[Z!/((Z-n)!n!)]e^{-\beta E_{n}}$.
 Here $E_n$ is the 
  electrostatic energy  
  of $n$ counterions  condensed onto the surface of a polyion
  and can be obtained through the charging process, 
\be
\label{/Zeta}
\beta E_{n}= -\beta q^2 n\int_{0}^{1} \frac{Z-n\lambda}{Da} 
d\lambda = -\frac{Zn-n^2/2}{T^{\ast}} \; .  
\ee

The minimization of the total free energy $f=f_{EN}+f_{PC}+f_{PP}+f_{CC}$
leads to the law of 
  mass action 
 $\mu_0+n\mu_f=\mu_n$. 
This is  a set of $Z$ coupled nonlinear algebraic equations.
We were able to solve these iteratively, starting 
 with a uniform distribution of clusters.
A sample of the  distribution obtained is presented 
in the inset of Fig.~1. The average cluster
size is $\langle n\rangle =\sum_{}^{} n\rho_n/\rho_p$,
and the average cluster charge is $\langle Z_{\rm eff}\rangle= Z-\langle n\rangle$.
We note that the width of the distribution remains quite
narrow and is not very sensitive to the variations in density or
temperature. This suggests that the polydispersivity in cluster
sizes is not very important and can
be replaced by {\it one}\/ characteristic
cluster size. In this case the theory becomes extremely simple,
since the free energy will be an
explicit function of the number of
bound counterions $n_B$, since $\rho_n=\rho_p \delta_{n,n_B}$. 
The thermodynamically
stable cluster, $\bar{n}$, will be the one for which
the free energy, $f(Z,n_B)$, attains a minimum, 
$f(Z,\bar{n})= \min_{n_B} f(Z,n_B)$. 
Thus, we must solve only {\it one}\/
algebraic equation, instead of
$Z$ coupled ones. Indeed, as expected, the one-cluster approximation
is in excellent agreement with the full theory. 
In Fig.~1 we present the average cluster charge 
as a function of the polyion bare
charge, while in Fig.~2 the osmotic pressure inside
the polyelectrolyte solution is computed, which is a monotonically
increasing function of the density of polyions.

\begin{figure}[hb]
\begin{center}
\leavevmode
\epsfxsize=0.45\textwidth
\epsfbox[0 0 520 410]{"
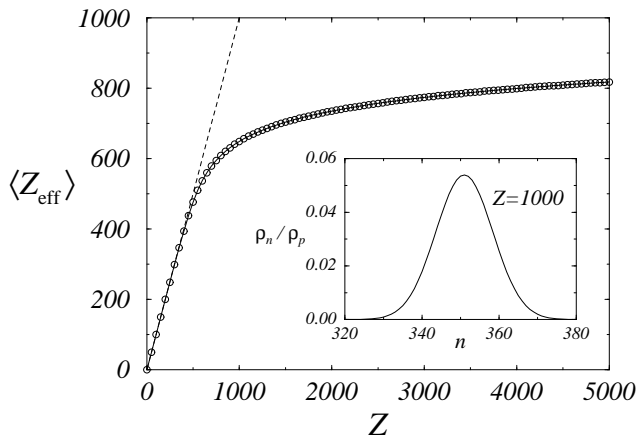"}
\end{center}
\begin{minipage}{0.48\textwidth}
\caption{Average cluster charge, $\langle Z_{\rm eff}\rangle$ 
as a function of
the bare  
charge $Z$ for a temperature $T^{\ast}=100$ and density     of
polyions
$\rho^{\ast}_{p}=3/(32\pi)$. The solid line is calculated using the
full theory (distribution of cluster sizes), while the
circles were obtained within  one-size  cluster
approximation. Inset depicts the  distribution of cluster sizes
for $Z=1000$.
The dashed line is the bare charge.}
\end{minipage}
\end{figure}

An interesting question that arises is, what happened to the
phase transition, so successfully predicted by the {\it DHBj}\/ 
theory in the case of symmetric electrolyte \cite{Fi93}?
The answer to this question is far from clear.
However, our derivation does shed some light on the
mechanism of the  disappearance of the phase
transition. The fundamental postulate
that only the 
counterions are polarized by the electric
field fluctuations,
 can be seem to 
lie behind the disappearance of the transition. Indeed if
this postulate is waved, 
so that both polyions and counterions can be 
polarized \cite{Br93}, 
it is a simple matter
to show
that the pure {\it DH}\/ theory 
predicts that the suspension will
undergo a phase separation when the temperature is
reduced below $T^{\ast}_{c}=Z/16$. 
The fundamental question of what is the
maximum charge asymmetry above which the
 polyelectrolyte solution will remain stable still needs
to be answered.

\begin{figure}[hb]
\begin{center}
\leavevmode
\epsfxsize=0.45\textwidth
\epsfbox[0 0 520 410]{"
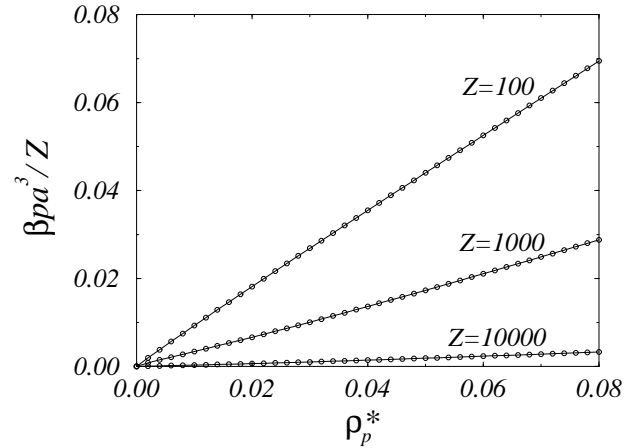"}
\end{center}
\begin{minipage}{0.48\textwidth}
\caption{Dependence of the renormalized
dimensionless pressure $\beta p a^3/Z$ on the density of
macroions $\rho^{\ast}_{p}$ for several values of $Z 
(100,1000,10000)$. 
The solid lines were calculated  using the full theory
(distribution), while the circles were obtained
using one-size cluster approximation.
}
\end{minipage}
\end{figure}

\section*{Acknowledgments}

We are grateful to Prof.~Pierre Turq and
to Prof.~Lesser Blum   for having introduced
us to the subject of charged colloids. Discussions and correspondence
with   Profs.~L. Belloni, M. E. Fisher, 
J.-P. Hansen and G. Stell have been greatly appreciated.
This work was supported in part by CNPq -- Conselho Nacional de
Desenvolvimento Cient\'{\i}fico e Tecnol\'ogico and FINEP --
Financiadora de Estudos e Projetos, Brazil.

\end{multicols}

\begin{thebibliography}{40}
\bibitem{Text} In the spirit of {\it DH} theory we assume that the
polyion has the same dielectric constant as the medium, see
Ref.~[14(d,e)].
\end{thebibliography}
\end{document}